\documentclass[aps,prl,twocolumn,showpacs,showkeys,groupedaddress,floatfix,amsmath,amssymb]{revtex4}
\usepackage{graphicx}
\usepackage{dcolumn}
\usepackage{bm}

\begin{document}


\title{Strong influence of a magnetic layer on the critical current of Nb
bridge in finite magnetic fields due to surface barrier effect}

\author{D. Y. Vodolazov, B. A. Gribkov, A. Yu. Klimov, V. V. Rogov and S. N. Vdovichev}

\affiliation{Institute for Physics of Microstructures, Russian
Academy of Sciences, 603950, Nizhny Novgorod, GSP-105, Russia}

\date{\today}

\begin{abstract}

We measured the critical current of the bilayer Nb/Co in the
applied magnetic field. When the magnetic field was tilted to the
axis which was perpendicular to the plane of the bilayer we
observed a large difference in critical currents flowing in
opposite directions. We found that the largest critical current of
the bilayer exceeded the critical current of the superconductor
without Co layer in a wide range of the tilted magnetic fields.
The theory which takes into account the surface barrier effect for
vortex entry and magnetic field of the magnetic layer gave a
quantitative explanation of our experimental results.

\end{abstract}

\maketitle

 The last decade has seen a large activity (see for example
\cite{Majoros,Glowacki,Genenko,Horvat,Touitou,Vodolazov1,Morelle,Jooss,Alamgir,Gomory1,Gomory2})
in studying the superconductor-ferromagnetic systems in which the
magnetic field of the magnet can be used for enhancing  the
critical current of the superconductor at zero and finite magnetic
fields. The main idea of these works is to use the ferromagnetic
sheath to shield magnetically the whole superconductor or the part
of it from the applied magnetic field and/or from the self-induced
field. It was experimentally proved that this method could lead to
improve the critical current at finite magnetic fields in
MgB$_2$/Fe wires \cite{Horvat} and BSCCO film covered by Ni
\cite{Alamgir} and to enhance the critical current in hybrid Nb/Co
structure in the parallel magnetic field \cite{Vodolazov1} as well
as in Al film placed close to the Co/Pt stripe in a perpendicular
magnetic field \cite{Morelle} and to decrease  ac losses in BSCCO
superconductors partially covered by Ni \cite{Gomory2}.

This work studies the dependence of the critical current ($I_c$)
of the Nb/Co bilayer on the direction and value of the applied
magnetic field and compares it with $I_c$ of the superconducting
bridge with a removed ferromagnetic layer. We prove that in our
sample both bulk pinning and a strong surface barrier for vortex
entrance exist. We demonstrate that in such superconductors the
magnetic layer has much stronger influence on its transport
properties than in superconductors with only bulk pinning
\cite{Alamgir,Horvat,Jooss}. The effect is mainly connected with a
strong dependence of the width of the vortex free region (which
exists in superconductors with a surface barrier
\cite{Kupriyanov,Benkraouda,Plourde}) on the local magnetic field.
We observe a big difference in the value of the critical current
depending on the direction of the current flow and we demonstrate
that the largest critical current of a bilayer is larger than the
critical current of the superconductor without magnetic layer in a
wide range of the magnetic fields.
\begin{figure}[hbtp]
\includegraphics[width=0.45\textwidth]{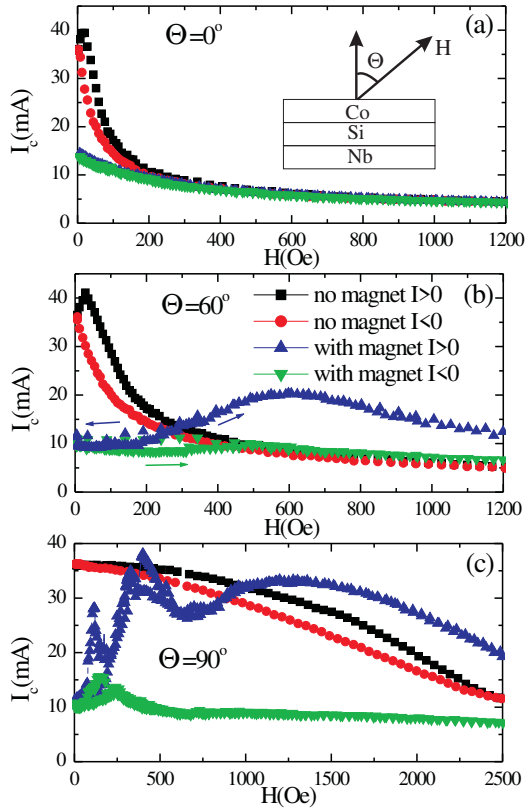}
\caption{(Color online)Dependence of the critical currents of
bilayer and superconducting bridge without magnetic layer on
applied magnetic field for three directions of H: a) magnetic
field is perpendicular to plane of the bilayer/bridge, b) magnetic
filed is tilted on $60^o$ to the axis which is perpendicular to
bilayer/bridge; c) magnetic field is parallel to bilayer/bridge.
In the insert we show the sketch of our sample and field
direction.}
\end{figure}

The samples were fabricated in one process from Nb/Si/Co
multilayer by Ar etching process in photoresist mask. The Nb
bridge was fabricated by a reactive magnetron spattering and it
had the critical temperature about 9.2 K. Ferromagnetic Co layer
and Si layer as dielectric interlayer were prepared by magnetron
spattering. The thicknesses of Nb, Si and Co layers were about 100
nm and the width and the length of the fabricated bridge were 2
and 6 $\mu m$, respectively. The magnetic state of the Co layer
was monitored by a Solver scanning probe microscope at room
temperature in the 'flying' mode. The saturation magnetization of
the Co layer was about 1200 G at 4.2 K. To measure the transport
properties of superconducting Nb bridge (without a magnet) the
ferromagnetic Co layer was removed in a weak solution of an acid.
The measurements were performed at the temperature T=4.2 K by the
standard four probe method.

In Fig. 1a-b we present the measured critical current of bilayer
and superconducting bridge without a magnetic layer for three
directions of the applied magnetic field. When the magnetic field
is perpendicular to the superconducting bridge the presence of the
ferromagnet leads to suppression of the critical current in all
magnetic fields (see Fig. 1a). The effect of Co layer is rather
different when the applied magnetic field is tilted to the axis
which is perpendicular to the plane of the bilayer (see Fig.
1b-c). First of all there exists a large difference in critical
currents flowing in opposite directions at $H \lesssim 2 kOe$.
Secondly, in low magnetic fields the value of the critical current
depends not only on the direction of the current flow but also on
the history. It is smaller when H is swept up and $I_c$ is larger
when H is swept down. Thirdly the largest critical current of the
bilayer is larger than the critical current of superconductor
without a magnet in a wide range of applied magnetic fields.

Let us to interpret our results using the model of the surface
barrier (SB) for vortex entry/exit from the superconductor and
magnetization of the ferromagnet by the applied magnetic field.
According to SB effect the dependence $I_c(H_{\bot})$ of the
superconducting film should be linear in low magnetic fields
\cite{Kupriyanov,Benkraouda,Plourde}. From the slope of a linear
dependence $I_c(H_{\bot})$ we extracted London penetration length
$\lambda \simeq 120 nm$ using the procedure of Ref.
\cite{Plourde}. The important parameter in the surface barrier
model is the critical current density at the edge of the
superconductor $j_s$ when the surface barrier is suppressed and
vortices can enter the superconductor. It should be equal to
depairing current density for the defect free superconductor
\cite{Kupriyanov,Plourde} and for real superconductors with
surface defects $j_s$ differs from sample to sample and can be
different for opposite edges of the superconducting bridge. In the
latter case it provides the difference in the value of the
critical current flowing in opposite directions
\cite{Plourde,Vodolazov2} in the finite magnetic field. Using the
dependence of $\lambda$ and the coherence length $\xi$ on mean
free path $\l$ in 'dirty' limit \cite{Tinkham} we estimated $\l
\simeq 5.5 nm$ and $\xi(4.2K)\simeq 14 nm$ for our sample (in
'pure' Nb $\xi\simeq\lambda\simeq 41 nm$). For these values the
depairing current density $j_{dep}\simeq j_{GL}=\sqrt{4/27}
c\Phi_0/8\pi^2\lambda^2\xi^2=4.9 \cdot 10^7 A/cm^2$ ($j_{GL}$ is a
critical current density in phenomenological Ginzburg-Landau
theory). From the comparison of the surface barrier model
($I_c(0)=j_s Wd\sqrt{\lambda^2/(Wd/2\pi+\lambda^2)}$ - see Ref.
\cite{Plourde}) and our maximal value of the critical current
$I_c(16Oe)\simeq 41 mA$ we find $j_s\simeq 0.72 j_{GL}$. If we
take $I_c(0)\simeq 36 mA$ we obtain a little bit smaller value
$j_s \simeq 0.64 j_{GL}$.
\begin{figure}[hbtp]
\includegraphics[width=0.45\textwidth]{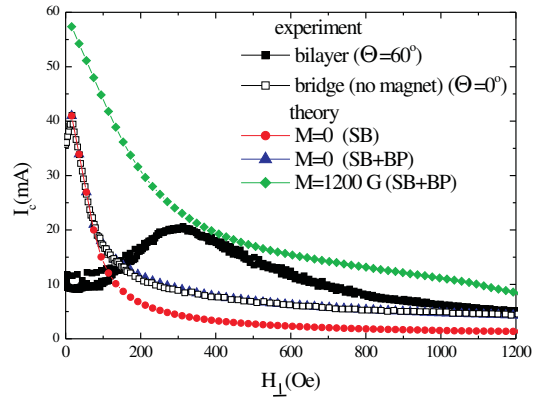}
\caption{(Color online) The experimental critical currents of
superconducting bridge without magnetic layer placed in
perpendicular magnetic field (empty squares) and bilayer (black
squares) placed in tilted $\Theta=60^o$ magnetic field as function
of $H_{\bot}$. Triangles, circles and rhombs are theoretical
values of critical current for our model system with zero and
finite magnetization in presence of surface barrier (SB) effect
and bulk pinning (BP).}
\end{figure}

To calculate the critical current of bilayer and bridge without a
magnet in an external magnetic field we used the same model as in
Ref. \cite{Vodolazov1}. To find the dependence $I_c(H_{\bot})$ we
numerically solved Eq. (1) of Ref. \cite{Vodolazov1} with
conditions: i) maximal current density in superconductor cannot
exceed $j_s$; ii) in the region occupied by vortices the current
density is equal to zero \cite{Kupriyanov,Plourde} or pinning
current density $j_p(H_{\bot})$. In numerical calculations we used
the parameters of our sample and neglected the influence of the
parallel magnetic field.

In Fig. 2 we plot dependencies $I_c(H_{\bot})$ found from the
experiment with a superconducting bridge with removed magnetic
layer in a perpendicular magnetic field and our theoretical
results. One can see that the theory which incorporates both
surface barrier effect and bulk pinning (BP) gives a better
agreement with the experiment than the theory with SB effect only.
We were able to extract the dependence $j_p(H_{\bot})$ from the
experiment with the bilayer in a perpendicular magnetic field
because the domain structure of the demagnetized magnet
efficiently suppresses the surface barrier effect (there are
regions along the bridge where magnetic fields of the domain and
transport current are summed and it effectively suppresses the
surface barrier for vortex entry). Fitting with expression
$j_p(H_{\bot})=j_{p0}/(1+(H_{\bot}/H_0))^{\beta}$ gives us
$j_{p0}=10^7 A/cm^2$, $H_0=128 Oe$ and $\beta=0.6$.

In Fig. 2 we also plot the calculated dependence $I_c(H_{\bot})$
at $M=1200 G$ (experimental value of saturation magnetization of
our Co layer) and experimental results. We see that in theory the
presence of the magnetized magnet leads to enhancing the critical
current in all magnetic fields. We neglect in our calculations the
influence of the  parallel magnetic field and it provides the
quantitative discrepancy between theory and experiment at large
fields where $H_{||}>500 Oe$ (compare Fig. 1b and 1c). Because in
our model we supposed that the magnet is in-plane magnetized with
constant magnetization M we were not able to describe the
dependence $I_c(H)$ of bilayer in low magnetic fields, where the
magnet has a domain structure (we checked it at room temperature
by MFM measurements). The increase of $I_c$ in low magnetic fields
may be explained by growing magnetization of initially
demagnetized magnet. It is clear that the magnetization may not
coincide with sweeping up and down magnetic field. It is the
reason for the hysteresis in critical currents observed in low
magnetic fields $H \lesssim 600 Oe$ in the experiment.

We can give a simple interpretation of enhancing the critical
current in a superconducting bridge in the presence of a magnetic
layer. In Fig. 3 we plot the calculated current density and
magnetic field distributions in superconducting film with and
without magnetized magnet at H$_{\bot}$=1 kOe. One can see that in
the region where the external magnetic field is partially or fully
compensated by the magnetic field of the magnet the current
density is much larger than the pinning current density and hence
there is no vortices. The width of a vortex free region with
increasing H decays much faster than the pinning current density
(compare Eq. (5) in Ref. \cite{Plourde} and our fitting expression
for $j_p(H_{\bot})$). When the magnetic field of the magnet
compensates the perpendicular component of the external magnetic
field in some part of the superconductor (see Fig. 3) the vortex
free region considerably expands and the critical current is
enhanced. The magnetic field of the magnet in the other part of
the superconductor does not influence the critical current much
because of a relatively weak dependence of $j_p(H_{\bot})$.
\begin{figure}[hbtp]
\includegraphics[width=0.45\textwidth]{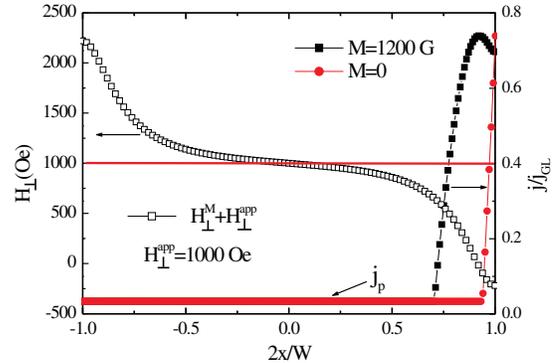}
\caption{(Color online)Modification of the current density
distribution in superconducting film placed in applied magnetic
field $H^{app}_{\bot}$=1 kOe at $I=I_c$ due to presence of the
magnetic field of fully magnetized magnet with M=1200 G (theory).}
\end{figure}

If the applied magnetic field is parallel to the bilayer
($\Theta=90^o$), we also observe the critical current enhancement
in large magnetic fields, but the reason for this effect is
slightly different from the discussed above. At our parameters the
parallel component of the magnetic field created by fully
magnetized magnet is equal to about 480 Oe in the superconductor.
Therefore the parallel magnetic field in superconducting film will
be 480 Oe smaller than the applied field (in fields large enough
to magnetize the magnet). It shifts the curve $I_c(H)$ to higher
fields. In tilted magnetic fields in addition the perpendicular
magnetic field is compensated in some part of the superconductor
and it affects the critical current stronger than the compensation
of the only parallel magnetic field.

In our previous work \cite{Vodolazov1} we used narrow and long
magnet to affect the critical current of the superconducting
bridge. In a demagnetized state it did not have a domain structure
(see Fig. 6 in Ref. \cite{Vodolazov1}) and did not suppress $I_c$
in zero magnetic field. But because it was much narrower than the
superconductor its influence was relatively weak. In present work
we covered the whole superconducting bridge by cobalt and it
resulted in much stronger influence on $I_c$ but as an edge effect
the domain structure appears at H=0 which suppresses $I_c$.
Apparently, to solve this problem one should use the ferromagnet
of the same length as the superconductor and  split it to series
of parallel stripes. It should favorite the magnetization of the
ferromagnetic stripes along the superconductor and do not affect
the critical current at H=0.

The work was supported by the Russian Foundation for Basic
Research and by the Fundamental Program "Quantum Macrophysics" of
the Russian Academy of Sciences. D.Y.V. acknowledges support from
Dynasty Foundation. S.N.V acknowledges support from Joint Program
'Basic Research and Higher Education'.

\end{document}